\documentclass[twocolumn,showpacs,preprintnumbers,amsmath,amssymb]{revtex4}
\usepackage{graphicx}  %for pdflatex use .jpg plots
\usepackage{dcolumn}% Align table columns on decimal point
\usepackage{bm}% bold math
\usepackage{latexsym} \usepackage{amssymb}

\begin{document} 
\hyphenation{dis-tri-bu-tion dis-tri-bu-tions 
oli-go-nu-cleo-tide Arab-i-dop-sis 
Meth-a-no-co-ccus jann-a-schii Tre-po-nema pall-i-dum Vib-rio chol-e-rae
Haem-o-phi-lus infl-uen-zae Chla-my-dia mur-i-da-rum du-pli-cat-ion 
re-pli-cat-ion chro-mo-some Ar-chaeo-glo-bus Stre-pto-co-ccus pneu-mo-niae
Clos-tri-dium ace-to-bu-tyli-cum}
%%%% code for the organisms
%\def\Atha{{\it Arabidopsis thaliana}}  \def\atha{{\it A. thaliana}}
\def\Aaeo{{\it Aquifex aeolicus}}  \def\aaeo{{\it A. aeolicus}}
\def\Aful{{\it Archaeoglobus fulgidus}}    \def\aful{{\it A. fulgidus}}  
\def\Aper{{\it Aeropyrum pernix}}  \def\aper{{\it A. pernix}} 
\def\Atum{{\it Agrobacterium tumefaciens}}  \def\atum{{\it A. tumefaciens}} 
\def\Bbur{{\it Borrelia burgdorferi}}  \def\bbur{{\it B. burgdorferi}}  
\def\Bhal{{\it Bacillus halodurans}} \def\bhal{{\it B. halodurans}} 
\def\Bmel{{\it Brucella melitensis}}   \def\Bmel{{\it B. melitensis}}   
\def\Bsub{{\it Bacillus subtilis}} \def\bsub{{\it B. subtilis}} 
\def\Busp{{\it Buchnera sp. APS}}   \def\busp{{\it B. sp.}}   
\def\Baph{{\it Buchnera aphidicola}}   \def\baph{{\it B. aphidicola}}   
\def\Cjej{{\it Campylobacter jejuni}} \def\cjej{{\it C. jejuni}} 
\def\Ccre{{\it Caulobacter crescentus}} \def\ccre{{\it C. crescentus}} 
\def\Cvib{{\it Caulobacter vibrioides}}  \def\cvib{{\it C. vibrioides}} 
\def\Clim{{\it Chlorobium limicola}} \def\clim{{\it Ch. limicola}} 
\def\Cmur{{\it Chlamydia muridarum}}  \def\cmur{{\it C. muridarum}}  
\def\Cace{{\it Clostridium acetobutylicum}}  \def\cace{{\it C. acetobutylicum}}
\def\Cper{{\it Clostridium perfringens}}  \def\cper{{\it C. perfringens}}   
\def\Cpne{{\it Chlamydia pneumoniae}}  \def\cpne{{\it Ch. pneumoniae}}  
\def\Ctra{{\it Chlamydia trachomatis}} \def\ctra{{\it Ch. trachomatis}} 
\def\Cglu{{\it Corynebacterium glutamicum}} \def\cglu{{\it C. glutamicum}}  
\def\Drad{{\it Deinococcus radiopugans}} \def\drad{{\it D. radiopugans}} 
\def\Ecol{{\it Escherichia coli}} \def\ecol{{\it E. coli}} 
\def\ecolK12{{\it E. coli K12}}  \def\ecolH7{{\it E. coli 0157:H7}}
\def\Fhep{{\it Flavobacterium heparinum}} \def\fhep{{\it F. heparinum}} 
\def\Fnuc{{\it Fusobacterium nucleatum}} \def\fnuc{{\it F. nucleatum}} 
\def\Gmax{{\it Glycine max}}  \def\gmax{{\it G. max}}    % Soybean
\def\Hasp{{\it Halobacterium sp.}}  \def\hasp{{\it H. sp.}} 
\def\Haur{{\it Herpetosiphon aurantiacus}} \def\haur{{\it H. aurantiacus}} 
\def\Hinf{{\it Haemophilus influenzae}} \def\hinf{{\it H. influenzae}} 
\def\Hpyl{{\it Helicbacter pylori}}  \def\hpyl{{\it H. pylori}}  
\def\Hvol{{\it Halobacterium volcanii}}  \def\hvol{{\it H. volcanii}}  
\def\Llac{{\it Lactococcus lactis}}  \def\llac{{\it L. lactis}}   
\def\Lmon{{\it Listeria monocytogenes}} \def\lmon{{\it L. monocytogenes}} 
\def\Mlot{{\it Mesorhizobium loti}}  \def\mlot{{\it M. loti}}  
\def\Mfer{{\it Methanothermus fervidus}} \def\mfer{{\it M. fervidus}} 
\def\Mjan{{\it Methanococcus jannaschii}} \def\mjan{{\it M. jannaschii}} 
\def\Mthe{{\it Methanobacterium thermoautotrophicum}} 
                     \def\mthe{{\it M. thermoautotrophicum}}  
\def\Mgen{{\it Mycoplasma genitalium}} \def\mgen{{\it M. genitalium}} 
\def\Mpen{{\it Mycoplasma penetrans}} \def\mpen{{\it M. penetrans}} 
\def\Mpne{{\it Mycoplasma pneumoniae}} \def\mpne{{\it M. pneumoniae}} 
\def\Mpul{{\it Mycoplasma pulmonis}} \def\mpul{{\it M. pulmonis}} 
\def\Mlep{{\it Mycobacterium leprae}} \def\mlep{{\it M. leprae}} 
\def\Mtub{{\it Mycobacterium tuberculosis}} \def\mtub{{\it M. tuberculosis}} 
\def\Nmen{{\it Neisseria meningitidis}} \def\nmen{{\it N. meningitidis}} 
\def\Neis{{\it Neisseria}}
\def\Nost{{\it Nostoc sp.}} \def\nost{{\it N. sp.}} % sp. PCC 7120 
\def\Paby{{\it Pyrococcus abyssi}}  \def\paby{{\it P. abyssi}}   
\def\Paero{{\it Pyrobaculum aerophilum}}  \def\paero{{\it P. aerophilum}}   
\def\Paeru{{\it Pseudomonas aeruginosa}} \def\paeru{{\it P. aeruginosa}} 
\def\Pfur{{\it Pyrococcus furiosus}}  \def\pfur{{\it P. furiosus}}   
\def\Phor{{\it Pyrococcus horikoshii}} \def\phor{{\it P. horikoshii}} 
\def\Pmul{{\it Pasteurella multocida}} \def\pmul{{\it P. multocida}} 
\def\Rnor{{\it Rattus norvegicus}} \def\rnor{{\it R. norvegicus}} % rat
\def\Rcon{{\it Rickettsia conorii}} \def\rcon{{\it R. conorii}} 
\def\Rpro{{\it Rickettsia prowazekii}} \def\rpro{{\it R. prowazekii}} 
\def\Rsol{{\it Ralstonia solanacearum}} \def\rsol{{\it R. solanacearum}} 
\def\Saur{{\it Staphylococcus aureus}} \def\saur{{\it S. aureus}} 
\def\Sent{{\it Salmonella enterica}} \def\sent{{\it S. enterica}} 
\def\Smel{{\it Sinorhizobium meliloti}}  \def\smel{{\it S. meliloti}}  
\def\Smel{{\it Sinorhizobium meliloti}}  \def\smel{{\it S. meliloti}}  
\def\Sfle{{\it Shigella flexneri}} \def\sfle{{\it S. flexneri}}  
\def\Sone{{\it Shewanella oneidensis}} \def\sone{{\it S. oneidensis}}  
\def\Spne{{\it Streptococcus pneumoniae}} \def\spne{{\it S. pneumoniae}}  
\def\Spyo{{\it Streptococcus pyogenes}} \def\spyo{{\it S. pyogenes}} 
\def\Scoe{{\it Streptomyces coelicolor}} \def\scoe{{\it S. coelicolor}} 
\def\Save{{\it  Streptomyces avermitilis}} \def\save{{\it S. avermitilis}}
\def\Ssol{{\it Sulfolobus solfataricus}} \def\ssol{{\it S. solfataricus}} 
\def\Stok{{\it Sulfolobus tokodaii}} \def\stok{{\it S. tokodaii}}  
\def\Stub{{\it Solanum tuberosum}} \def\stub{{\it S. tuberosum}} % white potato
\def\Styp{{\it Salmonella typhimurium LT2}} \def\styp{{\it S. typhimurium}} 
\def\Sent{{\it Salmonella enterica}} \def\sent{{\it S. enterica}} 
\def\Syne{{\it Synechococcus sp.}} \def\syne{{\it S. sp.}}
\def\Taci{{\it Thermoplasma acidophilum}}  \def\taci{{\it T. acidophilum}}   
\def\Tmar{{\it Thermotoga maritima}} \def\tmar{{\it T. maritima}} 
\def\Tpal{{\it Treponema pallidum}} \def\tpal{{\it T. pallidum}} 
\def\Tten{{\it Thermoprotues tenax}}  \def\tten{{\it T. tenax}}  
\def\Tvol{{\it Thermoplasma volcanium}} \def\tvol{{\it T. volcanium}}  
\def\Telo{{\it Thermosynechococcus elongatus}} \def\telo{{\it T. elongatus}}  
\def\Uure{{\it Ureaplasma urealyticum}} \def\uure{{\it U. urealyticum}} 
\def\Vcho{{\it Vibrio cholerae}} \def\vcho{{\it V. cholerae}} 
\def\Xfas{{\it Xylella fastidiosa}} \def\xfas{{\it X. fastidiosa}} 
\def\Ypes{{\it Yersinia pestis}}  \def\ypes{{\it Y. pestis}}
%%%%%%%%%%%%%% Eukaryotes %%%%%%%%%%%%%%%%%
\def\Agam{{\it Anopheles gambiae}}\def\agam{{\it A. gambiae}} %malaria vector
\def\Atha{{\it Arabidopsis thaliana}} \def\atha{{\it A. thaliana}} %mustard
\def\Asat{{\it Avena sativa}}\def\asat{{\it A. sativa}} %oat
\def\Cele{{\it Caenorhabditis elegans}} \def\cele{{\it C. elegans}}%nematode
\def\Drer{{\it Danio rerio}} \def\drer{{\it D. rerio}}% zebrafish
\def\Dmel{{\it Drosophila melanogaster}} \def\dmel{{\it D. melanogaster}} 
\def\Ecun{{\it Encephalitozoon cuniculi}}\def\ecun{{\it E. cuniculi}}
\def\Gmax{{\it Glycine max}}\def\gmax{{\it G. max}}% (soybean) 
\def\Gthe{{\it Guillardia theta}}\def\gthe{{\it G. theta}}% nucleomorph
\def\Hsap{{\it Homo sapiens}} \def\hsap{{\it H. sapiens}} 
\def\Lesc{{\it Lycopersicon esculentum}}\def\lesc{{\it L. esculentum}}%tomato
\def\Mmus{{\it Mus musculus}} \def\mmus{{\it M. musculus}} % House mouse
\def\Esat{{\it Oryza sativa}}\def\esat{{\it O. sativa}} %rice
\def\Pfal{{\it Plasmodium falciparum}}\def\pfal{{\it P. falciparum}} %malaria
\def\Plas{{\it Plasmodium}}  %malaria
\def\Rnor{{\it Rattus norvegicus}} \def\rnor{{\it R. norvegicus}} % rat
\def\Scer{{\it Saccharomyces cerevisiae}}\def\scer{{\it S. cerevisiae}} %yeast
\def\Spom{{\it Schizosaccharomyces pombe}} \def\spom{{\it S. pombe}}
\def\Taes{{\it Triticum aestivum}}\def\taes{{\it T. aestivum}} %bread wheat
\def\Amay{{\it Zea mays}}\def\amay{{\it Z. mays}} %corn
%%%%%%%%%%%%%%% common names %%%%%%%%%%%%%%%%%%%%%%%%%%
\def\bread_wheat{{\it T. aestivum}}
\def\corn{{\it Z. mays}} 
\def\fission_yeast{{\it S. pombe}}
\def\fly{{\it D. melanogaster}}
\def\human{{\it H. sapiens}}
\def\malaria{{\it P. falciparum}}
\def\malariavector{{\it A. gambiae}}
\def\mouse{{\it M. musculus}}
\def\mustard{{\it A. thaliana}}
\def\oat{{\it A. sativa}} 
\def\rat{{\it R. norvegicus}}
\def\rice{{\it O. sativa}}
\def\soybean{{\it G. max}}
\def\tomato{{\it L. esculentum}}
\def\worm{{\it C. elegans}}
\def\yeast{{\it S. cerevisiae}} 
\def\zebrafish{{\it D. rerio}}
%%%%%%%%%%%%%% additional abbreviations %%%%%%%%%%%%%%%%%
\def\ecols{{\it E. col.}}  \def\mjans{{\it M. jan.}} 
\def\cmurs{{\it C. mur.}}  \def\tpals{{\it T.pal}}
%%%%%%%%% End of genome names %%%%%%%%%%%%%%
\def\newpar{{\par\noindent}}
\def\newsec#1{{\sn\bf{#1}}}
\def\skipaline{{\vskip 12pt plus 1pt}}
\def\skiphafline{{\vskip 6pt plus 1pt}}
\def\qline{{\vskip 3pt plus 1pt}} \def\hfline{{\skiphafline}}
\def\sslash{{\slash\hskip -5pt}}
\def\slasha{{\sslash a}}
\def\mn{{\medskip\par\noindent}}
\def\bn{{\bigskip\par\noindent}}
\def\sn{{\smallskip\par\noindent}}
\def\sig{{\sigma}}
\def\olig{{oligonucleotide}}  \def\oligs{{\olig s}}
\def\oligm{{oligomer}}  \def\oligms{{\oligm s}}
\def\kmer{{$k$-mer}} \def\kmers{{$k$-mers}}
\def\dist{{distribution}}  \def\dists{{\dist s}}
\def\kdist{{$k$-distribution}}  \def\kdists{{\kdist s}}
\def\lroot{{L_{r}}}
\def\std{{std}}  \def\stds{{stds}}
\def\olig{{oligonucleotide}}  \def\oligs{{\olig s}}
\def\oligm{{oligomer}}  \def\oligms{{\oligm s}}
\def\kmer{{$k$-mer}}          \def\kmers{{\kmer s}}
\def\sigs{{$\sigma_S$}}   \def\sigl{{$\sigma_L$}}  \def\sig{{$\sigma$}} 
\def\ets{{$\eta_S$}}   \def\etl{{$\eta_L$}} 
\def\dist{{distribution}}  \def\dists{{\dist s}}
\def\foc{{occurrence frequency}}  \def\focs{{occurrence frequencies}}
\def\rmd{{root-mean-deviation}}
\def\mean#1{{\bar{#1}}}
\def\std{{s.d.}}  \def\stds{{s.d.'s}}
\def\mset{{$m$-set}} \def\msets{{$m$-sets}}  
\def\MM{{\cal M}}  \def\MMs{{\MM_{\sigma}}} 
\def\MMi{{\MM_R}}  \def\MMiz{{\MM^{(0)}_R}}
\def\Rzero{{R^{(0)}}}
\def\SS{{\cal S}} \def\QQ{{\cal Q}}    \def\FF{{\cal F}}
\def\ie{{\it i.e.}}
\def\etal{{\it et al.}}
\def\nrep{{$n$-replica}}   \def\nreps{{$n$-replicas}}   
\def\mmul{{$n$-multiple}}  \def\mmuls{{$n$-multiples}}
\def\qr{{quasireplica}} \def\qrs{{quasireplicas}} \def\Qrs{{Quasireplicas}}
\def\qrn{{quasireplication}}
\def\kspec{{$k$-spectrum}}   \def\kspecs{{$k$-spectra}}
\def\kband{{$k$-band}}   \def\kbands{{$k$-bands}}
\def\lbar{{\bar l}}
\def\ApT{{(A+T)}}  \def\CpG{{(C+G)}}
\def\refeq#1{{Eq.~(\ref{#1})}}
\def\reffg#1{{Fig.~\ref{#1}}}
\def\reftb#1{{Table~\ref{#1}}}
\def\peqhalf{{$p$$\approx$0.5}} \def\pnehalf{{$p$$\ne$0.5}}
\def\bgeq{\begin{equation}} \def\edeq{\end{equation}}
\def\SI{{Shannon information}} 
\def\FFk{{\FF^{\{k\}}}} \def\FFkm{{\FF_m^{\{k\}}}}
\def\FFQk{{\FF_\QQ^{\{k\}}}} \def\FFqk{{\FF_\calq^{\{k\}}}}
\def\calq{{\it q}}
\def\SSk{{\SS^{\{k\}}}} \def\SSkm{{\SS^{\{k,m\}}}}
\def\pahalf{{$p$$\approx$0.5}}
%%%%%%%%%%%%% Fonts %%%%%%%%%%%%%%%%%%%%%%%%%%%%%
\newfont{\ssflarge}{cmssi17 scaled 1320}  % Title
\newfont{\sfimed}{cmssi12 scaled 1200}    % Authors
\newfont{\sfi}{cmssi9 scaled 1200}        % Affiliation
\newfont{\sfb}{cmssdc10 scaled 1000}      % ABSTRACT
\newfont{\sfl}{cmssdc10 scaled 1320}      % SECTIONHEAD
\newfont{\nob}{cmb10 scaled 1200}         % Subtsectionhead

%%%%%%%%%%%%%%%%%%%%%%%%%%%%%%%%%%%%%%%%%%%%%%%%%

\title
{\bf Divergence and Shannon information in genomes}

\author{Hong-Da Chen$^1$, Chang-Heng Chang$^1$, 
Li-Ching Hsieh$^4$ and Hoong-Chien Lee$^{1-3}$}
\affiliation{
$^1$Department of Physics, $^2$Department of Life Science,  
and $^3$Center for Complex Systems, 
National Central University, Chungli, Taiwan 320, ROC\\
$^4$Institute of Information Science and Genomics Research Center, 
Academia Sinica, Taipei, Taiwan 115, ROC
}

\date{Received September 23, 2004; Revised \today}

\begin{abstract}
Shannon information (SI) and its special case, divergence, are defined
for a DNA sequence in terms of probabilities of chemical words in the
sequence and are computed for a set of complete genomes highly diverse
in length and composition.  We find the following: SI (but not
divergence) is inversely proportional to sequence length for a random
sequence but is length-independent for genomes; the genomic SI is
always greater and, for shorter words and longer sequences, hundreds
to thousands times greater than the SI in a random sequence whose
length and composition match those of the genome; genomic SIs appear
to have word-length dependent universal values. The universality is
inferred to be an evolution footprint of a universal mode for genome
growth.
\end{abstract}
\pacs{PACS number: 87.10.+e, 89.70.+c, 87.14.Gg, 87.23.Kg, 02.50.-r}
\maketitle

Shannon entropy \cite{Shannon48} has been used in almost every  
field concerned with information, including the 
analysis of DNA and protein sequences 
\cite{Gatlin72,Clote00,Bernaola00,Grosse02}.  
In the field of comparative genomics however it seems not
to have found any systematic application. The high heterogeneity of complete 
genomes in length, base composition and percentage of coding 
regions may make comparison based on Shannon entropy problematic.    
Here we show that by simple and appropriate definition of a quantity we 
call {\it Shannon information} (SI), difficulties associated with these 
issues are surmounted.  The SI are applied to characterize 
\dist\ of \foc\ (FD) of \kmers, or words of a $k$ 
chemical letters, in complete genomes. 
We present a simple relation 
between the SI and the relative spectral width of an FD,   
a relation that furnishes an 
intuitive understanding between SI and information in a sequence. 
We show that in spite of their high heterogeneity, 
the SIs in complete genomes can be represented by a set of 
genome independent universal lengths. 

\newsec{Divergence and Shannon information}. 
Consider a set of occurrence frequencies,
$\FF=\{f_i|\sum_{i=1}^{\tau}f_i=L\}\equiv \{f_i|\tau,L\}$, 
for $\tau$ types of events, where $f_i/L$ is the 
probability of event $i$.   
The Shannon entropy, or uncertainty \cite{Shannon48}, for $\FF$ is
%\begin{equation}
%H(\FF) = -\sum\nolimits_i (f_i/L) \ln (f_i/L)
$H(\FF) = -\sum_i (f_i/L) \ln (f_i/L)$. 
%\label{e:Sh_entropy}
%\end{equation}
The quantity attains its 
maximum value $H_{max}$=$\ln\tau$ 
when all $f_i$ are equal to the mean frequency $\bar f$=$\tau^{-1}L$.  
In $H$, Shannon was concerned with the fidelity of messages as 
they are transported through communication devices. Here we 
are interested in the information in $\FF$ itself.
There is a general notion that information in a system increases 
with a decrease in uncertainty, hence we identify the quantity, 
\begin{equation}
D(\FF) \equiv \ln\tau - H(\FF) = 
L^{-1} \sum\nolimits_i f_i \ln (f_i/\bar f)
\label{e:divergence}
\end{equation}
called {\it divergence} by Gatlin \cite{Gatlin72}, as the zeroth order 
SI in $\FF$. 
Shannon called the ratio $D(\FF)/H_{max}$ redundancy.  
%and reported that in ``ordinary 
%English, not considering statistical structure over greater than 
%about eight letters, (it) is roughly 50\%'' \cite{Shannon48}.  
%We will see that for genomes the 
%redundancy will be far less than 50\% in all cases. 
%Sometimes the association between $D$ and information appears  
%counterintuitive.  An extreme case is when 
%$f_i$ is equal to unity for one $i$ and to zero for all others.  
%Then $H(\FF)$ vanishes and $D(\FF)$ acquires its maximum 
%value $H_{max}$, and it would be better to say the system has 
%minimum uncertainty than to say that it has maximum information.  
%Such extreme cases do not occur in genomes. 

Suppose the set $\FF=$$\{\FF_m|m=1,2,\cdots\}$ is composed of subsets, 
$\FF_m=\{f_i|\tau_m,L_m\}$, each having its own distinct 
types of events $\tau_m$, 
total number of events $L_m$ and mean frequency $\bar f_m$=$L_m/\tau_m$, 
where $\sum_m\tau_m$=$\tau$ and $\sum_m L_m$=$L$.  The divergence of each 
subset is $D(\FF_m)$=$\sum'_i f_i/L_m \ln (f_i/\bar f_m)$, where the 
summation is restricted to those $f_i$'s in the subset $\FF_m$.  
We define the SI carried by $\FF$ to be the weighed average of the 
divergences in the subsets:
\bgeq
R(\FF)\equiv\sum_m (L_m/L) D(\FF_m) 
%={\sum_{m,i}}' \frac{f_i}{L} \ln (f_i/\bar f_m) 
=L^{-1}{\sum_{m,i}}' f_i \ln (f_i/\bar f_m) 
\label{e:Shaninf}
\edeq
%$D(\FF)$ approaches $R(\FF)$ when all  $\bar f_m$ approach $\bar f$. 

\newsec{Frequency \dist\ in a DNA sequence}.
We view a single strand of DNA and
as a linear text written in the four chemical letters, A, C,
G and T representing the four kinds of nucleotides.  
Empirically genomes are invariably within a 
few percent of being compositionally 
self-complementary and, for the present study,
%meaning that on a {\it single} 
%strand the numbers of A's and T's are close to 
%being equal, as are the numbers of C's and G's.  For the present study, 
it suffices to characterize 
the base composition of a genome by a single number, $p$, 
the combined probability of (A+T).  From now on the 
term {\it profile} of a sequence will refer to the $p$ value and the length 
$L$ of the sequence.  We will use the SI in random sequences as benchmarks 
for the SI in genomes.  A random sequence having the profile of a 
genome is said to be a {\it random match} for the genome.

For a DNA sequence of length $L$ we denote by $\FFk$ the 
set $\{f_i|\tau=4^k,L\}_k$, 
where $f_i$ is the \foc\ of the $i^{th}$ overlapping 
$k$-letter word, or \kmer\ \cite{Hao00}.  
%The frequencies are obtained by sliding a 
%window of width $k$ across the genome, one letter at a time, and 
%recording the number of times each \kmer\ is seen through the 
%window \cite{Hao00}.   
For $\FFk$ the number of event types is $\tau$=$4^k$ and the mean 
frequency is $\bar f$=$4^{-k} L$.  
Given $\FFk$ we can construct a {\it \kspec} where $n_f$, the number of 
\kmers\ occurring with frequency $f$, is given as a functiion of $f$.  
For simplicity we also call $\FFk$ a \kspec.
%Note that in this instance 
%we are not keeping track of the \foc\ of each \kmer. 
%The number of event types is now $\tau$=$4^k$, and we have the sum rules 
%$\sum_i 1$=$\sum_f n_f$=$4^k$ and $\sum_i f_i$= $\sum_f f n_f$=$L$.   
%To simplify our language we will also call $\FFk$ a \kspec.  

\begin{figure}[ht!] 
\vspace{-0.4cm}
\begin{center}
\includegraphics[width=3.0in,height=2.3in]
%{/Users/hclee/hclee/Sace_Ran30702M_RRC303005_05l_tmp6}
{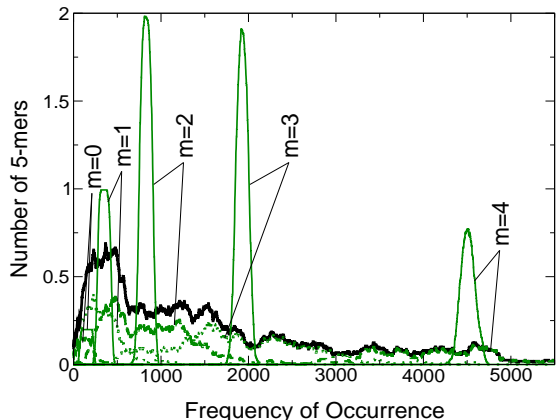}
\end{center}
\vspace{-0.8cm}
\caption{\baselineskip=8pt\label{f:Cace_5} \scriptsize\sf{
The 5-pectra of the genome of \cace\ (black) ($p$$\approx$0.7)
and that of its random match (green or gray), normalized to that of 
a 2 Mb sequence.  
The random-match spectrum is composed of 
sharply peaking subspectra $\FFkm$, $m$= 0 to 4. The center of the $m$=5 
subspectrum is off scale at 10,500.  
Ordinates $n_f$ are not integers because (for better viewing only) 
large fluctuations in the spectra 
are smoothed out by averaging over a range of 21 frequencies.}}
\vspace{-0.3cm}
\end{figure}

The $\FFk$ of a genomes is always composed of 
$k$+1 subsets $\FFkm$, called {\it \msets}, $m$=1 to $k$, where $\FFkm$ 
is the subset of \kmers\ with $m$ \ApT's.  
When a genome does not have \pahalf, and most genomes are of this 
type, then the mean frequencies ${\bar f}_m$ of the subsets $\FFkm$ 
are distinct: ${\bar f}_m(p)$=$\mean{f} 2^k p^m (1-p)^{k-m}$. 
\reffg{f:Cace_5} \cite{GenBank} 
shows the 5-spectra of a $p$=0.691 genome, \Cace, and 
its random match. The spectrum from the random match (sharp peaks in 
gray (or green)) is 
composed of five nonoverlapping subspectra (a sixth one is out of range), 
one for each $\FF_m^{\{5\}}$. The corresponding 
subspectra (dash and dotted gray curves) underlying the entire  
genomic spectrum (black curve) are also shown; they are seen to have  
overlapping widths. 
 
The types and number of \kmers\ in $\FFkm$ are $\tau_m$=$2^k (k,m)$, 
and $L_m$=$\tau_m\bar f_m$, respectively, where $(k,m)$ is the binomial 
coefficient. 
As in \refeq{e:Shaninf} we 
define the SI in $\FFk$ as the weighed mean of the divergences in $\FFkm$: 
$R(\FFk)$=${\sum}'_{m,i} {f_i/L} \ln ({f_i/\bar f_m})$. 
When $p$ approaches 0.5 all $\bar f_m$ tend to $\bar f$ 
and $R(\FFk)$ approaches $D(\FFk)$. Note that $R(\FFk)$ and $D(\FFk)$ 
are invariant under the replacement $p$ $\to$ 1-$p$. 

\newsec{Relative spectral width}.
In \refeq{e:divergence}, by expanding $f_i$ around $\bar f$ in 
the logarithm we obtain for a unimodal $\FF$ the series
%Given a spectrum $\FF$ with mean $\bar f$ and standard deviation 
%from the mean $\Delta$, we define $\sigma$$\equiv$$\Delta/{\bar f}$ 
%as the {\it relative spectral width} (RSW) of $\FF$.  In statistics, $\sigma$ 
%given in percentages is known as the coefficient 
%of variation of $\FF$.  The divergence is related to the RSW of $\FF$  
%by \cite{Lee_unpub}
\bgeq
D(\FF) = \sum_{n=2} {(-1)^n\over n(n+1)}
\left\langle\left({f_i-\bar f\over \bar f}\right)^n\right\rangle
\approx{\Delta^2\over 2 {\bar f}^2}  
%+ {\cal O}\left(\sigma^3\right) 
%\qquad ({\rm unimodal}\ \FF)
\label{e:Sh_info_sigma}
\edeq
where $\Delta$ is the standard deviation in $\FF$, or 
$D(\FF)\approx \sigma^2/2$, where $\sigma$ 
is the {\it relative spectral width} (RSW) of $\FF$. 
This relation is  useful in at least two respects. 
First, it gives one a heuristic understanding why the divergence 
of and information in a (unimodal) spectrum are connected.  
When the spectrum is narrow, 
there is little information because all the \kmers\ occur 
with almost equal frequency. 
%Conversely, so long as $\sigma$$<$1, the broader the spectrum the 
%higher the potential information content.  
Because of the generally substantiated notion that 
highly overrepresented words in a genome are more likely to have 
biological meaning than words represented with an average 
frequency \cite{Smith95,Karlin96,Helden98,Bussemaker00}, 
the obverse is also true, namely, there is more information 
when the spectrum in broad.  
Second, \refeq{e:Sh_info_sigma} gives an accurate estimate of 
the divergence in the $\FFkm$ (and $\FFk$) of a random sequence. 
Provided $\bar f_m$ is much greater than one, 
$\FFkm$ is proportional to a Poisson \dist\ with mean $\bar f_m$ 
\cite{Xie02,Hsieh03}, so that 
$\Delta_{ran}^2$$\approx$$b_k \bar f_m$ and 
$\sigma_{ran}^2$$\approx$$b_k/\bar f_m$.   Hence 
$D(\FFkm)\approx b_k/2\bar f_m = b_k \tau_m/2L_m$ for a random sequence. 
Then from \refeq{e:Shaninf} we have a result for the SI in the $\FFk$ 
of any random match, independent of $p$: 
\bgeq
R(\FFk)\approx b_k \tau/2 L \qquad({\rm Random\ sequence})
\label{e:Sh_random}
\edeq
The combinatorial factor $b_k$ should be 1-$\tau^{-1}$ 
for a true random sequence; a semi-empirical value  
$1-1/2^{k-1}$ is used here because a random match is not fully 
random: it is made to be approximately compositionally self-complementary 
(as most genomes are) and its $p$ value is fixed.  
That the SI in a random sequence diminishes as $1/L$ with increasing $L$  
is connected to what is known as the large-system rule: the square of 
the RSW of a \dist\ associated with a random system is inversely proportional 
to the size of the system.  We will see that the SI in genomes does 
not follow this rule.

\begin{table} 
\vspace{-0.3cm}
\caption{\baselineskip=8pt\label{t:Sh_ecoli}\footnotesize\sf{ 
Shannon entropy $H$ and divergence $D$ in units of $\ln 2$ in the \kspecs\ 
of the genome sequence of \ecolH7\ and its random match.}} 
\begin{ruledtabular}
\begin{tabular}{ccccccc}
&\multicolumn{3}{c}{\underline{\phantom{HHH} Random match \phantom{HHH}}}&
\multicolumn{3}{c}{\underline{\phantom{HHH}\ecolH7\phantom{HHH}}}\\
$k$& $H$ & $D$ &$b_k 4^k/2L$ & $H$&$D$&$\sigma^2/2$\\
\hline
 2 & 4.0000& 5.12E-7& 1.06E-6& 3.986& 1.39E-2& 1.39E-2\\
 3 & 6.0000& 6.76E-6& 6.26E-6& 5.953& 4.71E-2& 4.41E-2\\
 4 & 8.0000& 3.03E-5& 2.92E-5& 7.908& 9.16E-2& 8.65E-2\\
 5 & 9.9999& 1.28E-4& 1.25E-4& 9.854& 1.46E-1& 1.42E-1\\
 6 & 11.999& 5.23E-4& 5.18E-4& 11.79& 2.07E-1& 2.11E-1\\
 7 & 13.998& 2.12E-3& 2.10E-3& 13.72& 2.73E-1& 2.90E-1\\
 8 & 15.991& 8.56E-3& 8.48E-3& 15.65& 3.49E-1& 3.87E-1\\
 9 & 17.965& 3.45E-2& 3.42E-2& 17.54& 4.54E-1& 5.22E-1\\
 10 & 19.858& 1.42E-1& 1.37E-1& 19.34& 6.58E-1& 7.70E-1\\
\end{tabular}
\end{ruledtabular}
\vspace{-0.4cm}
\end{table}

\newsec{Shannon information in a \pahalf\ genome}.
The genome of \ecolH7\ is 5.53 Mb long and has $p$=0.496. 
We therefore use \refeq{e:divergence} 
to compute its SI. 
The Shannon entropy and divergence (in units of $\ln 2$) in the \kspecs,
$k$=2 to 10, of the genome and its random match are given in 
\reftb{t:Sh_ecoli}.  We notice the following: 
({\it i}) For both sequences the Shannon entropy (columns 2 and 5)
is in every case  close to its maximum value, 2$k$.  
({\it ii}) For the random match the value computed using \refeq{e:divergence} 
(column 3) is in excellent agreement with the expected value (column 4). 
({\it iii}) For the smallest $k$'s, the divergence is a minuscule 
signal buried in a huge Shannon entropy background. 
({\it iv}) This tiny signal, when isolated, cleanly sets apart a 
genome from its random match:  the genomic divergence is greater than 
its random-match counterpart in all cases and, for the 
smaller $k$'s, many thousand times greater. 
({\it v}) \refeq{e:Sh_info_sigma} is verified (last column). 
% features will be shown to be generic for the SI  
%in all the complete genomes we have examined. 

\newsec{Difference between divergence and SI}. 
When $p$ is not close to 0.5 $D(\FFk)$ and $F(\FFk)$ differ.  Consider
the 5-spectra (not the subspectra) for the genome and its random match
shown in \reffg{f:Cace_5}.  Because the $\bar f_m$'s are spread
widely, their \dist\ play a dominant role in determining the width of
the spectra.  If we ignore the widths of the individual subspectra and
use \refeq{e:Sh_info_sigma} we get $D(\FFk)$$\approx$$D^{(0)}(k,p)$=$
0.5[(2^k(p^2 + (1-p)^2))^k -1]$ for both the genomic and random
spectra ($D^{(0)}(5,0.691)$=0.488, as compared to the actual values of
0.575 and 0.485 for the genomic and random 5-spectra in
\reffg{f:Cace_5}).  $D(\FFk)$ depends strongly on $p$, is independent
of $L$, disagrees with \refeq{e:Sh_random}, and is therefore not useful 
for estimating the SI of a random sequence.  
%If it were, we would be
%faced with a senseless result: a genome and its random match having
%vastly different SIs when
%\pahalf\ (\reftb{t:Sh_ecoli}) but having almost the same SI when
%$p$$\ne$0.5.

The monomer divergence and SI are quantities that are completely
determined by the base composition of a sequence and are independent
of its randomness.  Whereas $R(\FF^{\{1\}})$=0 always,
$D(\FF^{\{1\}})$ depends strongly on $p$ and vanishes only at
$p$=0.5. The genome of \Mgen\ has $p$=0.683 and $D(\FF^{\{1\}})$=0.0686 
(compared with $D^{(0)}$=0.670). 
The genome \Hinf\ has $p$=0.618 and $D(\FF^{\{1\}})$=0.0281
(0.0278).  Since these values say nothing about the randomness or
order of the sequence yet depend nontrivially on $p$, they convey {\it
less} information about the sequence than the value of $p$ alone.

\begin{table} [h!]
\vspace{-0.3cm}
\caption{\baselineskip=8pt\label{t:k6_m-set}\footnotesize\sf{ 
Divergence (\refeq{e:divergence}) in the \msets\ of the 6-spectra of three 
genomes and their random matches.}} 
\begin{ruledtabular}
\begin{tabular}{cccccc}
$m$ & $\bar f_m$ &  $L_m$ (kb) &  
$b_6/(2\bar f_m)$ & $D_{ran}$ & $D_{gen}$\\
\hline
\multicolumn{6}{l}{\vspace{3pt}\Paero, $p$=0.486, $L$= 2.22 Mb}\\
$0$ &  480 &  30.7 &  1.01E-3 &  8.54E-4 &  4.32E-2\\
1   &  500 &   192 &  9.69E-4 &  1.06E-3 &  6.88E-2\\
2   &  520 &   499 &  9.31E-4 &  9.58E-4 &  8.71E-2\\
3   &  541 &   692 &  8.96E-4 &  8.12E-4 & 1.33E-1\\
4   &  563 &   540 &  8.93E-4 &  8.80E-4 & 1.80E-1\\
5   &  586 &   225 &  8.60E-4 &  9.37E-4 & 1.58E-1\\
6   &  610 &  39.0 &  7.94E-4 &  7.65E-4 &  7.74E-2\\
\hline
\multicolumn{6}{l}{\vspace{3pt}\Cmur, $p$=0.597, $L$= 1.07 Mb}\\
$0$& 68.7 & 4.40 & 7.05E-3 & 4.87E-3 & 1.11E-1\\
1 &   103 & 39.5 & 4.70E-3 & 4.47E-3 & 1.16E-1\\
2 &   155 &  148 & 3.12E-3 & 3.18E-3 & 1.15E-1\\
3 &   232 &  297 & 2.08E-3 & 2.15E-3 & 1.07E-1\\
4 &   348 &  334 & 1.39E-3 & 1.49E-3 & 1.02E-1\\
5 &   521 &  200 & 9.29E-4 & 1.12E-3 & 1.20E-1\\
6 &   782 & 50.0 & 6.14E-4 & 7.36E-4& 1.52E-1\\
\hline
\multicolumn{6}{l}{\vspace{3pt}\Save, $p$=0.293, $L$= 9.03 Mb}\\
$0$&  89.0 &    5.70 &  5.44E-3 &  5.99E-3 &  1.73E-1\\
1 &    214 &    82.2 &  2.26E-3 &  2.27E-3 &  1.80E-1\\
2 &    519 &     498 &  9.33E-4 &  1.04E-3 &  4.21E-1\\
3 &   1252 &   1,602 &  3.87E-4 &  3.78E-4 &  2.59E-1\\
4 &   3024 &   2,903 &  1.60E-4 &  1.66E-4 &  1.33E-1\\
5 &   7300 &   2,803 &  6.63E-5 &  6.14E-5 &  6.22E-2\\
6 & 17,622 &   1,127 &  2.75E-5 &  2.87E-5 &  8.87E-2\\
\end{tabular}
\end{ruledtabular}
\vspace{-0.25cm}
\end{table}

\newsec{The general case}.
\reftb{t:k6_m-set} gives the divergences, $D_{gen}$ and $D_{ran}$, 
of the \msets\ in the 6-spectra 
of three sequences and their random matches, \paero, \cmur, and \save. 
These sequences span a wide range in profile.  Once again, 
$D_{ran}$ (column 5) is well approximated by its expected value (column 4) 
and is much less than $D_{gen}$.  $D_{gen}$ varies but does not 
exhibit a clear dependence on $L$.  
The 6-mer SI, computed according to \refeq{e:Shaninf}, for \ecolH7\ and 
the three genomes of \reftb{t:k6_m-set}
and their random matches are given in \reftb{t:k6_ave}. 
Again we have: $R_{ran}$ is well approximated  
by $b_k\tau/2L$, $R_{gen}$$>>$$R_{ran}$, and $R_{gen}$  
is independent of genome length.

\begin{table}[t!] 
\vspace{-0.35cm}
\caption{\baselineskip=8pt\label{t:k6_ave}\footnotesize\sf{ 
Shannon information (\refeq{e:Shaninf}) 
in the 6-spectra of the four genomes cited in 
Tables \ref{t:Sh_ecoli} and \ref{t:k6_m-set} and their random matches.}} 
\begin{ruledtabular}
\begin{tabular}{ccccccc} 
Genome&$p$ & $L$ (Mb) & $\bar f$ &  $R_{ran}$ & 
$R_{ran}/(b_6/2\bar f)$ & $R_{gen}$\\
\hline
{\it E. coli}&0.486&5.52&1348&3.59E-4&1.01&0.143\\
{\it P. aero.}&0.514&2.22&542&8.97E-4&0.995&0.127\\
{\it C. muri.}&0.597&1.07&261&1.83E-3&1.05&0.108\\
{\it S. aver.}&0.293&9.03&2203&2.20E-4&1.03&0.144\\
\end{tabular}
\end{ruledtabular}
\vspace{-0.25cm}
\end{table}
\begin{figure}[t!] 
%\vspace{-0.2cm}
\begin{center}
\includegraphics[width=3.2in,height=2.5in]
%{/Users/hclee/hclee/DNA/entropy/dat/Shaninf}
{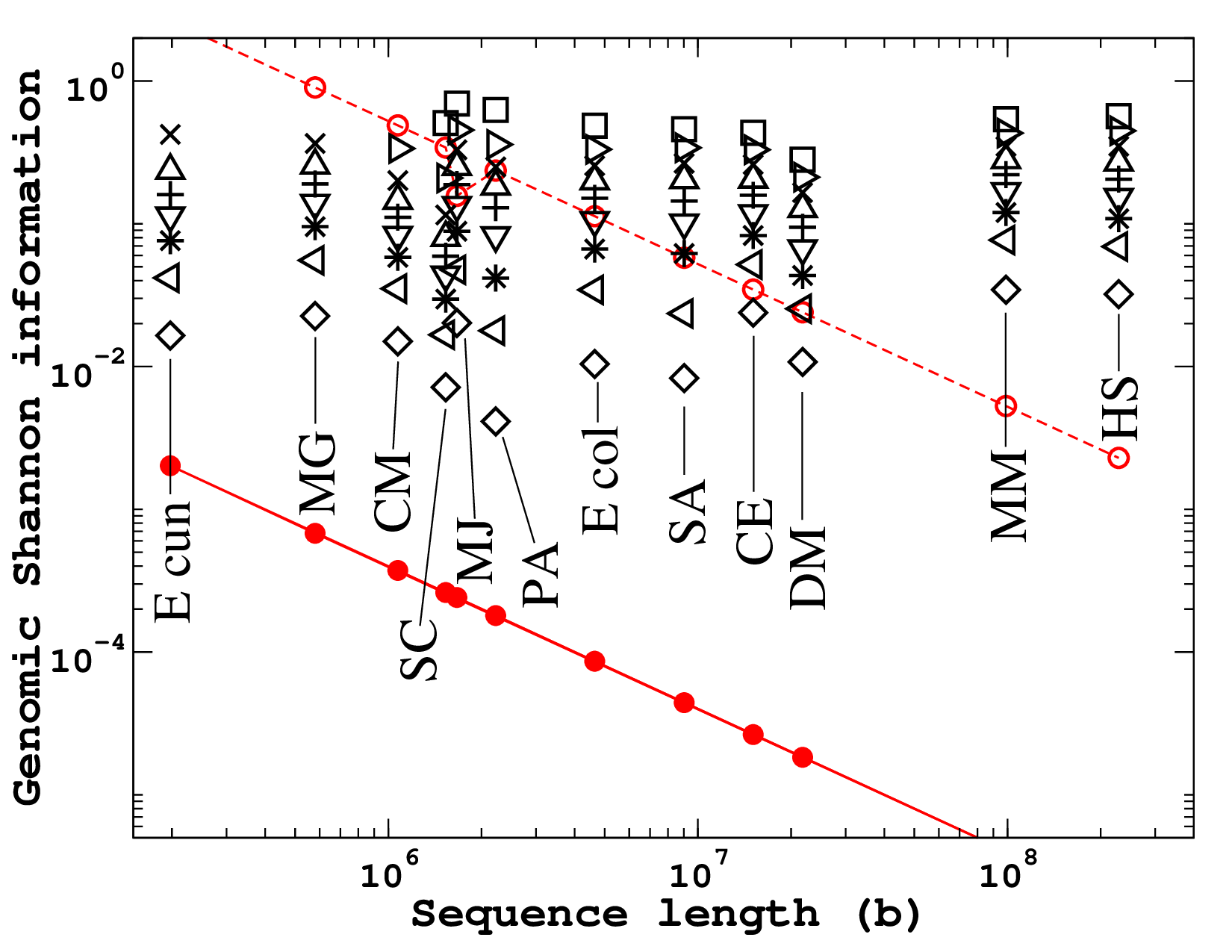}
\end{center}
\vspace{-0.8cm}
\caption{\baselineskip=8pt\label{f:twelve_genomes}{\scriptsize\sf{
SI in the \kspecs\ for $k$=2 ($\Diamond$), 3 ($\triangleleft$), 4 (*),
5 ($\triangledown$), 6 (+), 7 ($\triangle$), 8 ($\times$), 9
($\triangleright$), and 10 ($\Box$), of the genomes of twelve
organisms: E cun, \ecun\ ({\it Chromosome I}; length 0.198 Mb,
$p$=0.533); CM, \cmur; MG, \mgen\ (length 0.580 Mb, $p$=0.683); SC,
yeast (\yeast, $IV$; 1.53 Mb, 0.621); MJ, \mjan\ (1.66 Mb, 0.686); PA,
\paero; E col, \ecolK12\ (4.64 Mb, 0.492); SA, \save; CE, worm (\worm,
{\it Chr. I}; 15.1 Mb, 0.643); DM, fly (\fly, $X$; 21.8 Mb, 0.574);
MM, mouse (\mmus, $IX$; 98.9 Mb, 0.562); HS, human (\human, $I$; 228
Mb, 0.582). The lines give SI expected in random sequences for $k$=10
(open circles connected by dotted line) and 100 times the expected 
value for $k$=2 (solid circles connected by solid line).
}}}
\vspace{-0.4cm}
\end{figure}

A new feature is manifest in \reftb{t:k6_ave}: in spite 
of their greatly varying profiles the four genomes have almost the same SI 
in their 6-spectra.   Indeed, at least for word lengths up to 10 letters, 
we have found that 
SI varies little for complete genomes with vastly differing profiles.  
\reffg{f:twelve_genomes} shows SI 
as a function of sequence length in six prokaryotes 
and six eukaryotic chromosomes \cite{GenBank}.  In the plot, the symbols in 
each column, top to bottom, give the \kmer\ SI, 
$k$=10 ($\Box$), 9 ($\triangleright$),\dots, to 2 ($\Diamond$), 
of an organism.  
The \kmer\ SI of a sequence is ill-defined when the sequence length 
is less than twice $4^k$, therefore the SIs for $k$=9 and 10 in  
\ecun\ and for $k$=10 in \mgen\ are not given.  We have verified 
that the SI in different chromosomes of the same organism vary little. 
For comparison the solid and dashed lines in \reffg{f:twelve_genomes} connect 
the expected 2-mer SI ($\times$100) and 10-mer SI    
in the matching random sequences.  

\newsec{Genomes have universal SI}. 
With the exception of several 2-mer SIs, all genomic SIs lie 
within a horizontal band bounded by 0.01 and 0.9.  
(For technical reasons too lengthy to describe here but which will be explained 
elsewhere, the 2-mer SI has the largest fluctuations.) 
For a given $k$ the genomic variance in SI is only about a factor of two. 
In comparison the longest sequence - \human\ ($I$) at 228 Mb - 
is about 1,200 times longer than the shortest - \ecun\ $I$ at 0.198 Mb.  
In this context we view the genomic SI as having 
a genome independent but $k$-dependent universal value.  For each 
$k$ we express this universal value, $R_{uni}(k)$, in terms of the length, 
$L_r$, called the {\it equivalent root-sequence}, 
of a random sequence whose SI is equal to $R_{uni}$: 
$L_r(k)$$\equiv$$4^k b_k/2R_{uni}(k)$.  Then the genomic results in 
\reffg{f:twelve_genomes} are summarized by 
the empirical relation $\ln(L_r(k))$=$\alpha k+C$, where 
$\alpha$=1.01$\pm$0.06, $C$=3.80$\pm$0.50. 
%%%%%%%%%  Compare with ``Reduced spectral width result''%%%%
%If use log_10, then
%     log L_eff = a k + B
%     a~ 0.434,  B~ 1.60 +- 0.20
%   This last result is consistent w/ result in "Science" paper
$L_r$ is surprisingly short for small $k$ but grows rapidly with $k$:
$L_r(k)$$\approx$340 b, 15 kb and 1.1 Mb,  
for $k$=2, 6, and 10, respectively. 

Now, if a random root-sequence of length $L_r$ is replicated a finite 
number of times, then, provided $k$$<<$$L_r$, the \kmer\ SI in the product 
sequence is the same as that in the root-sequence. 
We therefore suugest the observed universality of SI in complete 
genomes is an evolution footprint  
produced by a universal mode of genome growth, a mode that 
began when the genomes were very short - less than 340 b - and in 
which segmental 
duplication played a major role \cite{Hsieh03,Lee_etal}.  
Work exploring the implication of this notion 
will be reported elsewhere.

This work is partly supported by grants 92-2112-M-008-040 and 
93-2311-B-008-006 from the National Science Council (ROC) to HCL. 

%\vspace{-10pt}

\end{document}